\documentclass[twocolumn, secnumaRabic, amssymb, nobibnotes, aps, showpacs,superscriptaddress]{revtex4-2}
\usepackage{graphicx,subfigure,amsmath}%
\usepackage{subfigure}
\usepackage{color}
\usepackage{bm}
\usepackage{epstopdf}
\usepackage[bookmarks=false]{hyperref}
\hypersetup{colorlinks=true,citecolor=blue,linkcolor=blue,urlcolor=blue,pdfstartview=FitH,bookmarksopen=true}
\usepackage{amsmath,amstext}
\usepackage{graphicx}
\usepackage{booktabs}
\usepackage{multirow}
\usepackage{float}
\begin{document}

\title{Optomechanically induced transparency on exceptional surfaces}

\author{Yang Pan}
\affiliation{School of Physics, Zhengzhou University, Zhengzhou 450001, China}
\author{H.-L. Zhang}
\affiliation{Academy for Quantum Science and Technology, Zhengzhou University of Light Industry, Zhengzhou 450002, China}
\affiliation{School of Electronics and Information, Zhengzhou University of Light Industry, Zhengzhou 450001, China}
\author{Y.-F. Jiao}
\affiliation{Academy for Quantum Science and Technology, Zhengzhou University of Light Industry, Zhengzhou 450002, China}
\affiliation{School of Electronics and Information, Zhengzhou University of Light Industry, Zhengzhou 450001, China}
\author{Y. Wang}
\affiliation{Academy for Quantum Science and Technology, Zhengzhou University of Light Industry, Zhengzhou 450002, China}
\affiliation{School of Electronics and Information, Zhengzhou University of Light Industry, Zhengzhou 450001, China}
\author{D.-Y. Wang}
\email{dywang@zzu.edu.cn}
\affiliation{School of Physics, Zhengzhou University, Zhengzhou 450001, China} 
\author{S.-L. Su}
\email{slsu@zzu.edu.cn}
\affiliation{School of Physics, Zhengzhou University, Zhengzhou 450001, China} 
\affiliation{Institute of Quantum Materials and Physics, Henan Academy of Sciences, Zhengzhou 450046, China}
\author{Hui Jing}
\email{jinghui73@foxmail.com}    
\affiliation{Key Laboratory of Low-Dimensional Quantum Structures and Quantum Control of Ministry of Education, Department of Physics and Synergetic Innovation Center for Quantum Effects and Applications, Hunan Normal University, Changsha, Hunan 410081, China}
\affiliation{Academy for Quantum Science and Technology, Zhengzhou University of Light Industry, Zhengzhou 450002, China}
\date{\today}

\begin{abstract}
Exceptional points (EPs) are singularities in non-Hermitian systems where the transmission spectrum exhibits significant variation at the phase transition point. Here, we propose a practical method for investigating the optomechanically induced transparency (OMIT) spectrum in a non-Hermitian cavity optomechanical system. By exploiting two kinds of exceptional surfaces (ESs) derived from the design of the waveguide structure, we study the transmission spectra and fast-slow light phenomena at various EPs on these ESs, and identify ES-protected transmission spectra that are robust to parameter fluctuations. Consequently, the system exhibits a complete reversal of both transmission and fast-slow light behavior at exceptional points (EPs) compared with those at non-EP (NP) when the phase transition occurs. Unlike previous studies that focus solely on an isolated EP, our approach enables the exploration of system characteristics with different EPs, providing greater flexibility for experimental design.
\end{abstract}

\maketitle

\section{INTRODUCTION}\label{sec1}
The exceptional point (EP)~\cite{ozdemirParityTimeSymmetry2019,miriExceptionalPointsOptics2019,PhysRevApplied.13.014070,heiss2012physics,heiss1990avoided}, as a critical point in non-Hermitian systems, plays a pivotal role in further improving measurement precision~\cite{PhysRevApplied.12.024002,chenExceptionalPointsEnhance2017c,mao2024exceptional,kononchuk2022exceptional}. Consequently, it is interesting to study the changes in the spectrum of the system at EPs for precise measurements~\cite{luOptomechanicallyInducedTransparency2018b,wang2020electromagnetically,xie2024exceptional}. Generally, EPs are the coalescence point of two or more eigenvectors of non-Hermitian Hamiltonian~\cite{heiss2004exceptional,fengNonHermitianPhotonicsBased2017,el-ganainyNonHermitianPhysicsPT2018}. The phenomenon of eigenvectors coalescing leads to a series of remarkable experimental findings~\cite{wiersig2014enhancing,pengLossinducedSuppressionRevival2014b,zhouObservationBulkFermi2018b,yangIdealWeylPoints2018b}. However, EPs are isolated in the parameter space in most studies~\cite{miriExceptionalPointsOptics2019,wang2021coherent}, so that they are highly sensitive to any unavoidable perturbation~\cite{li2024chip,li2023exceptional,wiersig2020prospects}, such as fabrication errors, experimental uncertainties, and environmental noises. To this end, researchers must precisely manipulate various parameters~\cite{peng2016chiral,Nanophotonics2022,PhysRevLett.113.053604,PhysRevA.92.013852,PhysRevA.99.043818} to ensure the attainment of EPs in non-Hermitian systems. For example, two nanopositioners are used to strictly control the positions of particles with nanometer-level precision to guarantee the presence of EPs~\cite{peng2016chiral,chenExceptionalPointsEnhance2017c}. In addition, it is essential to ensure that the two particles have extremely small but distinct differences~\cite{luOptomechanicallyInducedTransparency2018b,chenExceptionalPointsEnhance2017c}. Alternatively, in the context of parity time symmetric implementation~\cite{Nanophotonics2022,PhysRevLett.113.053604,PhysRevA.92.013852,PhysRevA.99.043818}, a delicate balance of systemic gain-loss alongside accurate control of coupling strength is necessary~\cite{hodaei2017enhanced}. To overcome these difficulties and enhance the system robustness, the concept of exceptional surface (ES) has been proposed~\cite{zhongSensingExceptionalSurfaces2019c,zhou2019exceptional,PhysRevA.105.L031501,Yang:21} and observed in numerous experiments~\cite{jiangExperimentalRealizationExceptional2023a,qinExperimentalRealizationSensitivity2021d,zhangExperimentalObservationExceptional2019a}. For example, a mirror is placed on one side of the tapered waveguide to achieve unidirectional coupling between the clockwise and counterclockwise modes, thereby establishing an ES~\cite{jiangExperimentalRealizationExceptional2023a}.

In parallel, cavity optomechanics (COM), exploring the interaction between optical fields and macroscopic mechanical resonators, has witnessed rapid development in recent years~\cite{metcalfeApplicationsCavityOptomechanics2014a,bekkerInjectionLockingElectrooptomechanical2017a,aspelmeyerCavityOptomechanics2014a,midoloNanooptoelectromechanicalSystems2018a} and yielded numerous research topics, such as quantum entanglement~\cite{liEnhancedEntanglementTwo2017,yanEntanglementOptimizationFiltered2019,liaoEntanglingTwoMacroscopic2014,stefanatosMaximisingOptomechanicalEntanglement2017,dengOptimizingOutputphotonEntanglement2016,vitaliOptomechanicalEntanglementMovable2007,liuPhasecontrolledAsymmetricOptomechanical2023}, non-classical mechanical states~\cite{mengMechanicalSqueezingFast2020,nationNonclassicalMechanicalStates2013,bergholmOptimalControlHybrid2019,renSinglephotonTransportMechanical2013}, quantum sensing~\cite{wang2024quantum,10.1063/5.0208107,zhao2020weak,zhang2024squeezing,PhysRevA.108.053514}, ground-state cooling of motion~\cite{chanLaserCoolingNanomechanical2011,teufelSidebandCoolingMicromechanical2011}, nonreciprocal optical transmission~\cite{xuOpticalNonreciprocityOptomechanical2015,heTransmissionNonreciprocityMutually2018}, normal mode splitting~\cite{huangNormalmodeSplittingCoupled2009,groblacherObservationStrongCoupling2009,dobrindtParametricNormalModeSplitting2008}, optoelectronic quantum transducers~\cite{lecocqMechanicallyMediatedMicrowave2016,bagciOpticalDetectionRadio2014,wangQuantumParametricAmplification2023}, phonon lasing~\cite{PhysRevLett.104.083901,PhysRevLett.113.053604}, mechanical squeezing~\cite{leiQuantumNondemolitionMeasurement2016}, etc. In particular, the experimental demonstration of OMIT has laid a foundation for diverse applications, offering an excellent platform for integrated optics and information processing. However, previous studies have mainly focused on NP regimes, the properties of OMIT spectrum on ESs have not been explored.

Here in this work, we study ES-protected OMIT in a non-Hermitian cavity optomechanical system, which consists of a whispering gallery mode (WGM) microresonator coupled with a single external nanoparticle to induce symmetric backscattering~\cite{qinExperimentalRealizationSensitivity2021d}. By incorporating a structural design that enables nonreciprocal transmission and asymmetric coupling~\cite{qinExperimentalRealizationSensitivity2021d,zhongSensingExceptionalSurfaces2019c}, we achieve a system capable of realizing two kinds of ES. The formation of ES results in the continuous connection of EPs in parameter space, significantly enhancing the robustness of the scheme to the parameter variation~\cite{zhongSensingExceptionalSurfaces2019c}. Leveraging these ESs, we study the OMIT spectra and their group delays at various EPs on two kinds of ESs. Additionally, we also generate non-EPs (NPs) by destroying the formation of ES, and compare them with the original EPs. In comparison with other schemes that involve changing complex optical gain~\cite{peng2016chiral} or precisely manipulating particle position and size~\cite{luOptomechanicallyInducedTransparency2018b,chenExceptionalPointsEnhance2017c}, our proposal offers a simpler and more cost-effective solution for investigating OMIT in non-Hermitian systems.

The remaining parts of the paper are organized as follows. In Sec.~\ref{sec2}, we present the Hamiltonian of the COM system in the rotating frame and then derive the conditions for generating two kinds of ESs. Subsequently, we calculate the transmission coefficient of the probe field. In Sec.~\ref{sec3}, we discuss the OMIT spectrum and group delay at different points. In Sec.~\ref{sec4}, we briefly discuss the feasibility of the experiment. The conclusion is presented in Sec.~\ref{sec5}.

\section{Theoretical model}\label{sec2}

\begin{figure}[htpb]
\centering
\includegraphics[width=0.49\textwidth]{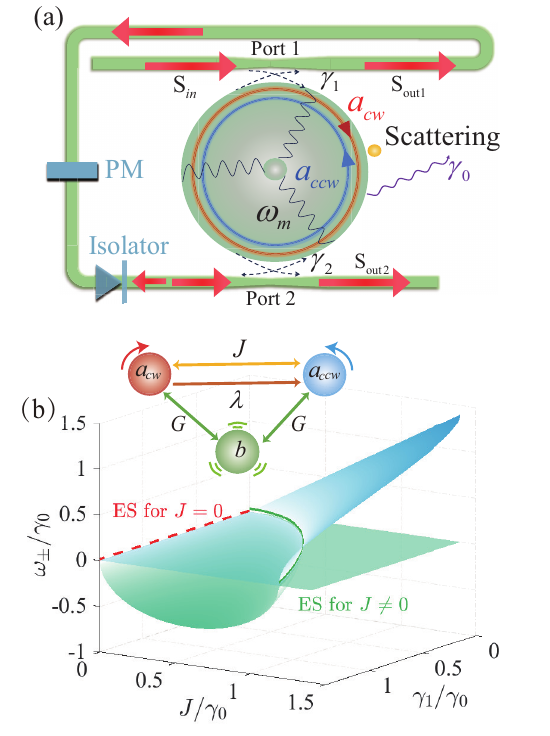}
\caption{(a) Schematic of the proposed system in experiments, where PM is the phase modulator. (b) Schematic of the non-Hermitian optomechanical system shown in the top, where the CW and CCW photon modes are coupled with the symmetric coupling strength $J$ induced by a nanoparticle and the asymmetric coupling $\lambda$ induced by the structural design. And the breathing phonon mode $b$ is coupled to the CW and CCW photon modes with the same optomechanical coupling coefficient $G$. Blow is the eigenvalues coalesce in our proposal. The dashed line represents the general condition when $J=0$. And the solid line corresponds the phase transition point with $J=t_0\sqrt{\gamma_1\gamma_2}$, which can form the ES in a higher dimensional parameter space.}\label{fig1}
\end{figure}

Inspired by a recent experiment, which demonstrates sensitivity enhancement in a non-Hermitian photonic system~\cite{qinExperimentalRealizationSensitivity2021d}, we consider a non-Hermitian optomechanical system comprising of a WGM microresonator and a tapered fiber, as shown in Fig.~\ref{fig1}. Due to manufacturing limitations, the WGM microresonator includes imperfections, which leads to backscattering of transmitted light~\cite{svela2020coherent}. For simplicity, we can attribute these backscattering to the coupling of a nanoparticle. At this point, the two transmitted modes of WGM microresonator, denoted as the clockwise (CW) mode $a_{\rm{cw}}$ and the counterclockwise (CCW) mode $a_{\rm{ccw}}$, are coupled to each other with the symmetric coupling strength $J$. To study OMIT in a non-Hermitian optomechanical system, the structure of the fiber is redesigned~\cite{zhongSensingExceptionalSurfaces2019c,qinExperimentalRealizationSensitivity2021d}, which results in the asymmetric one-way coupling between the two transmitted modes with effective coupling constant $\lambda$.

In our proposal, the microresonator, with resonance frequency $\omega_{0}$ and intrinsic decay rate $\gamma_{0}$, supports a mechanical mode with mechanical frequency $\omega_{m}$ and damping rate $\gamma_{m}$. The optomechanical coupling strength between the mechanical mode and both the optical transmitted modes is denoted as $G=\omega_0/R$~\cite{PhysRevLett.104.083901}, where $R$ is the radius of the microresonator. The system is driven by a strong pump field with frequency $\omega_{c}$ and a weak probe field with frequency $\omega_{p}$. The amplitudes of the pump field and probe field are $E_{c}=\sqrt{P_{c}/\hbar\omega_{c}}$, $E_{p}=\sqrt{P_{p}/\hbar\omega_{p}}$, where $P_{c}$ and $P_{p}$ is the laser power of the pump field and probe field. The Hamiltonian~\cite{metcalfeApplicationsCavityOptomechanics2014a,luOptomechanicallyInducedTransparency2018b,qinExperimentalRealizationSensitivity2021d} describing the total system can be written as ($\hbar=1$)

\begin{align}\label{equ1} 
		H & =H_{0} +H_{\rm{int}} +H_{\rm{cw}} +H_{\rm{ccw}},\\ 
		H_{0} & =\ \omega _{0} a_{\rm{cw}}^{\dagger } a_{\rm{cw}} +\omega _{0} a_{\rm{ccw}}^{\dagger } a_{\rm{ccw}}+\frac{p^{2}}{2m} +\frac{1}{2}m\omega _{m}^{2}x^{2} \nonumber,\\ 
		H_{\rm{int}} & =- Gx\left( a_{\rm{cw}}^{\dagger } a_{\rm{cw}} +a_{\rm{ccw}}^{\dagger } a_{\rm{ccw}}\right)+ J (a_{\rm{cw}}^{\dagger } a_{\rm{ccw}} + a_{\rm{ccw}}^{\dagger } a_{\rm{cw}})\nonumber,\\ 
		H_{\rm{cw}} & =i\sqrt{\gamma _{1}}\left[\left( E_{c}e^{-i\omega_{c} t} +E_{p} e^{-i\omega_{p} t}\right) a_{\rm{cw}}^{\dagger } -\text{H.c.}\right]\nonumber,\\ 
		H_{\rm{ccw}} & =i\sqrt{\gamma _{2}} \left[\left( t_{1} E_{c} e^{-i\omega_{c} t} +t_{2} E_{p} e^{-i\omega_{p} t} \right) a_{\rm{ccw}}^{\dagger }-\text{H.c.}\right]\nonumber\\ 
		&\quad+\lambda a_{\rm{cw}}a_{\rm{ccw}}^{\dagger } \nonumber.
\end{align}
Here, $\lambda$ is the asymmetric one-way coupling constant, which is described by $\lambda=i\sqrt{\gamma_1\gamma_2}t_3$~\cite{qinExperimentalRealizationSensitivity2021d}. $\gamma_{1}$ and $\gamma_{2}$ are the coupling coefficients between the fiber and the microresonator at Port 1 and Port 2, respectively. The displacement and momentum operators of the mechanical oscillator are denoted by $x$ and $p$. $m$ is the effective mass of the mechanical oscillator. $t_{j}=t_{0}e^{i\phi_{j}}$~\cite{zhongSensingExceptionalSurfaces2019c,qinExperimentalRealizationSensitivity2021d} is the transmission coefficient of different transmission light between Port 1 and Port 2 in the redesigned fiber, which is determined by the frequency of the transmission light and the loss coefficient of the fiber.

Since the single photon optomechanical coupling strength is negligible, we focus exclusively on the non-Hermitian properties of the optical subsystem~\cite{PhysRevA.107.033507,PhysRevApplied.12.024002}. We designate the corresponding eigenvalues of the non-Hermitian system as $\mathcal{E}_{\pm}=\omega_{\pm}-i\kappa_{\pm}$, which can be derived as
\begin{align}
\omega _{\pm} =\pm \sqrt{\alpha +\beta } \label{equ2},\\
\kappa _{\pm} =\pm \sqrt{\alpha -\beta }\label{equ3},
\end{align}	
where
\begin{align}
\alpha&=\frac{\sqrt{J^{4} +2J^{3} t_{0}\sqrt{\gamma _{1} \gamma _{2}}\sin( \phi _{3}) +J^{2} t_{0}^{2} \gamma _{1} \gamma _{2}}}{2}\label{equ4},\\
\beta&=\frac{J^{2} +Jt_{0}\sqrt{\gamma _{1} \gamma _{2}}\sin( \phi _{3})}{2}\label{equ5}.
\end{align}

For simplicity, the symmetric coupling $J$ between the two transmitted modes is considered to be a real number~\cite{chenExceptionalPointsEnhance2017c,qinExperimentalRealizationSensitivity2021d}. According to the theory of non-Hermitian physics, the eigenvalues of system would coalesce at the phase transition point. Here, $\omega_{+}=\omega_{-}$ and $ \kappa_{+}=\kappa_{-}$ mean that the eigenvalues of system coalesce when the system parameters $\left\{J,~t_{0},~\gamma_{1},~\gamma_{2},~\phi_{3}\right\}$ satisfy a specific relationship. At this time, the surface formed by the parameter relationship is called ES in non-Hermitian physics. For instance, when $J = 0$~\cite{qinExperimentalRealizationSensitivity2021d}, meaning the symmetric coupling between the two transmitted modes is nonexistent, the eigenvalues are $\omega_{\pm}=0$ and $\kappa_{\pm}=0$, as indicated by the red dashed line in Fig.~\ref{fig1}(b). The system is at EP with fixed by the parameters $\left\{\gamma_1,~\gamma_2,~t_0,~\phi_3\right\}$. The surface that unfolds in the parameter space is called ES. One EP can be distinguished from another through $\left\{\gamma_1,~\gamma_2,~t_0,~\phi_3\right\}$.   The other situation is when $J\neq0$, and the phase $\phi_{3}$ is tuned to $1.5\pi$~\cite{qinExperimentalRealizationSensitivity2021d}, Eqs. (\ref{equ2}) and (\ref{equ3}) can be simplified as
\begin{align}
\omega _{\pm} =\pm \sqrt{\frac{|J^{2} -J t_{0}\sqrt{ \gamma _{1} \gamma _{2} }|}{2} +\frac{J^{2} -Jt_{0}\sqrt{\gamma _{1} \gamma _{2}}}{2}}\label{equ6},\\
\kappa _{\pm} =\pm \sqrt{\frac{|J^{2} -J t_{0}\sqrt{ \gamma _{1} \gamma _{2} }|}{2} -\frac{J^{2} -Jt_{0}\sqrt{\gamma _{1} \gamma _{2}}}{2}}.\label{equ7}
\end{align}	
It is easy to observe that when the symmetric coupling satisfied the following relationship
\begin{equation}\label{equ8}
J=t_{0}\sqrt{\gamma_{1}\gamma_{2}},
\end{equation}
the two eigenvalues of the system are the same. And the result is shown in the green solid line of Fig.~\ref{fig1}(b). The point corresponding to the above relationship in Eq.~(\ref{equ8}) is the so-called phase transition point where the first-order quantum phase transition occurs in the non-Hermitian system. The surface that unfolds in the parameter space according to the relationship is also ES.

We have pointed out the two situations producing ES in the studied non-Hermitian system. Next, we will derive the transmission spectrum of the system and study the effect of ES on the transmission spectrum. In the rotating frame with the pump field frequency, $U=\exp{[i \omega_c(a^{\dagger}_{\rm{cw}}a_{\rm{cw}}+a^{\dagger}_{\rm{ccw}}a_{\rm{ccw}})t]}$, the dynamical quantum Langevin equations of the system can be given by
\begin{align}\label{equ9}
\dot{a}_{\rm{cw}} &=-( i\Delta -igx+\gamma ) a_{\rm{cw}} -iJa_{\rm{ccw}} +\sqrt{\gamma _{1}} E_{c} \nonumber\\
		&\quad+\sqrt{\gamma _{1}} E_{p} e^{-i\xi t} -\sqrt{\gamma _{1} \gamma _{2}} t_{3} a_{\rm{ccw}} \nonumber,\\
\dot{a}_{\rm{ccw}} &=-( i\Delta -igx+\gamma ) a_{\rm{ccw}} -iJa_{\rm{cw}}\sqrt{\gamma _{2}} t_{1} E_{c} \nonumber\\
		&\quad +\sqrt{\gamma _{2}} t_{2} E_{p} e^{-i\xi t}+\sqrt{\gamma _{1} \gamma _{2}} t_{3} a_{\rm{cw}}\nonumber,\\
\ddot{x} &=-\gamma _{m}\dot{x} -\omega _{m}^{2} x^{2} +\frac{g}{m}\left( a_{\rm{cw}}^{\dagger } a_{\rm{cw}} +a_{\rm{ccw}}^{\dagger } a_{\rm{ccw}}\right),
\end{align}
where $\xi=\omega_{p}-\omega_{c}$, $\Delta=\Delta_{a}+J$, $\Delta_{a}=\omega_{0}-\omega_{c}$, and $\gamma=(\gamma_{0}+\gamma_{1}+\gamma_{2})/2$. When the system reaches its steady state, all dynamical variables no longer change with time. Meanwhile, the steady-state values of the dynamical variables can be given by
\begin{align}\label{equ10}
\overline{a}_{\rm{cw}} &=\frac{E_{c}\left(\sqrt{\gamma _{1}}( i\Delta -ig\overline{x}+\gamma ) -t_{1}\sqrt{\gamma _{2}}\left( t_{3}\sqrt{\gamma _{1} \gamma _{2}} +iJ\right)\right)}{( i\Delta -ig\overline{x}+\gamma )^{2} +\gamma _{1} \gamma _{2} t_{3}^{2} +J^{2}}\nonumber,\\
\overline{a}_{\rm{ccw}} &=\frac{E_{c}\left( t_{1}\sqrt{\gamma _{2}}( i\Delta -ig\overline{x}+\gamma ) +\sqrt{\gamma _{1}}\left( t_{3}\sqrt{\gamma _{1} \gamma _{2}} -iJ\right)\right)}{( i\Delta -ig\overline{x}+\gamma )^{2} +\gamma _{1} \gamma _{2} t_{3}^{2} +J^{2}}\nonumber,\\
\overline{x}& =\frac{g}{m \omega _{m}^{2}}(\left\lvert \overline{a}_{\rm{cw}}\right\rvert^2+\left\lvert \overline{a}_{\rm{ccw}}\right\rvert^2).
\end{align}

To study the OMIT in this system, we calculate the optical transmission rate by linearizing the system operators, i.e., expanding the operator as the sum of its steady value and a small fluctuation $O(t)=\overline{O}+\delta O(t)$~\cite{metcalfeApplicationsCavityOptomechanics2014a,weisOptomechanicallyInducedTransparency2010e}. Specifically, the system operators are expanded as
\begin{align}
		x&=\overline{x} +\delta xe^{-i\xi t}+\delta x^{*}e^{i\xi t},\label{equ11}\\
		a_{j}&=\overline{a}_{j} +\delta a_{j}^{-} e^{-i\xi t} +\delta a_{j}^{+} e^{i\xi t}, \quad (j=\rm{cw},\rm{ccw}).\label{equ12}
\end{align}
Substituting Eqs.~(\ref{equ11}) and (\ref{equ12}) into Eq.~(\ref{equ9}), we can obtain
\begin{align}\label{equ13}
\chi ^{-1} \delta x-g\sum _{j={\rm{cw},\rm{ccw}}}^{{}} (\overline{a}_{j}^{*} \delta a_{j}^{-} +\overline{a}_{j} \delta a_{j}^{+*} )&=0\nonumber,\\
f_{1} \delta a_{{\rm{cw}}}^{-} +\left(t_{3}\sqrt{\gamma_{2}}+iJ\right) \delta a_{{\rm{ccw}}}^{-} -ig\overline{a}_{{\rm{cw}}} \delta x&=\sqrt{\gamma _{1}} E_{p}\nonumber,\\
f_{2} \delta a_{{\rm{cw}}}^{+*} +\left(t_{3}^{*}\sqrt{\gamma_{2}}-iJ^{*}\right)\delta a_{{\rm{cw}}}^{+*} +ig\overline{a}_{{\rm{cw}}}^{*} \delta x&=0,\\
f_{1} \delta a_{{\rm{ccw}}}^{-} -\left(t_{3}\sqrt{\gamma_{2}}-iJ\right) \delta a_{{\rm{cw}}}^{-}  -ig\overline{a}_{{\rm{ccw}}} \delta x&=t_{2}\sqrt{\gamma _{2}} E_{p}\nonumber,\\
f_{2} \delta a_{{\rm{ccw}}}^{+*} -\left(t_{3}^{*}\sqrt{\gamma_{2}}+iJ^{*}\right) \delta a_{{\rm{cw}}}^{+*} +ig\overline{a}_{{\rm{cw}}}^{*} \delta x&=0\nonumber,
\end{align}
where $f_{1,2}=\gamma-i\xi\pm i\left(\Delta-g\overline{x}\right)$ and $\chi^{-1}=m(\omega_{m}^{2}-\xi^{2}-i\xi\gamma_{m})$. The results of $\delta a^{-}_{{\rm{cw}}}$ and $\delta a_{\rm{ccw}}^{-}$ can be calculated by solving Eqs.~(\ref{equ13}) (See detailed results in Appendix~\ref{sec7}). Now, we can derive the output fields at Port 1 and Port 2 using the standard input-output relationships
\begin{align}\label{equ14}
s_{\rm{out2}}&=t_{2}s_{\rm{in}}-t_{3}\sqrt{\gamma_{1}}\delta a_{\rm{cw}}^{-}-\sqrt{\gamma_{2}}\delta a_{\rm{ccw}}^{-},
\end{align}
where $s_{\rm{in}}$ and $s_{\rm{out2}}$ are the input and output field operators, respectively. The transmission coefficient is defined by $t=s_{\rm{out2}}/s_{\rm{in}}$. Then, the transmission rate of the probe field in this system is obtained by
\begin{equation}\label{equ15}
T=\left\lvert t \right\rvert ^{2}=\left\lvert t_{2}-\frac{t_{3}\sqrt{\gamma_{1}}\delta a_{\rm{cw}}^{-}+\sqrt{\gamma_{2}}\delta a_{\rm{ccw}}^{-}}{E_{p}}\right\rvert ^{2}.
\end{equation}

In this section, we have discussed the appearance of the two kinds of ES (with $J=0$ or $J\neq0$) and derived the transmission spectrum of the system. It lays the groundwork for the forthcoming discussion regarding the effects of ES on OMIT spectrum and the group delay.

\section{Results and discussions}\label{sec3}

Before the discussions, we give the selected experimentally feasible parameter values based on previous reports~\cite{peng2014parity,luOptomechanicallyInducedTransparency2018b,qinExperimentalRealizationSensitivity2021d}, i.e., $R=34.5~\mathrm{\mu m}$, $\omega_{0}=193~\mathrm{THz}$, $\gamma_{0}=1~\mathrm{MHz}$, $m=50~\mathrm{ng}$, $\omega_{m}=147~\mathrm{MHz}$, $P_{c}=1~\mathrm{mW}$, and $\gamma_{m}=0.24~\mathrm{MHz}$. The other parameters in our proposal will be also selected based on the current experimental conditions.

\subsection{The first kind of exceptional surface}

\begin{figure}[htpb]
	\centering
	\includegraphics[width=0.48\textwidth]{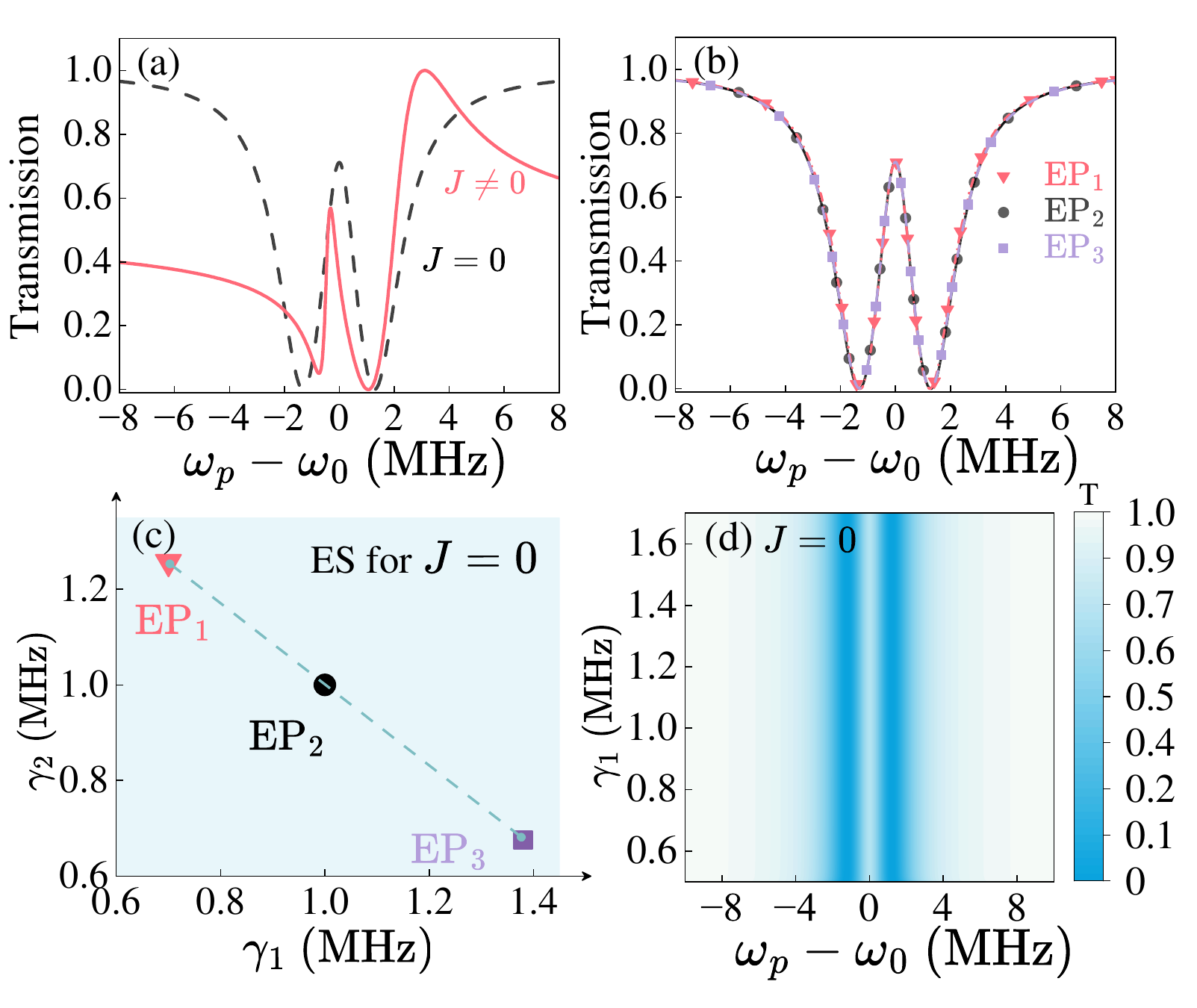}
	\caption{(a, b) The transmission spectra of probe light as a function of the optical detuning $\delta_p=\omega_p-\omega_0$. (c) The schematic diagram of the ES for $J=0$ spanned by $\gamma_1$ and $\gamma_2$, where $\rm{EP}_1$, $\rm{EP}_2$, and $\rm{EP}_3$ are three EPs on the same line with $\{\gamma_1,~\gamma_2\}=\{0.7~\rm{MHz},~1.26~\rm{MHz}\},~\{1~\rm{MHz},~1~\rm{MHz}\}$, and \{1.38~\rm{MHz},~0.68~\rm{MHz}\}, respectively. The equation of the line is $\gamma_2=f(\gamma_1)=-0.86\gamma_1+1.86$. (d) The transmission spectra on the dashed blue line in Fig.~\ref{fig2}(c).}\label{fig2}
\end{figure}

Figure~\ref{fig2}(a) presents the two kinds of OMIT spectra, each corresponding to the system being on a different ESs for $J=0$ or $J\neq0$. The black dashed line represents the transmission spectrum when the symmetric coupling disappears $J=0$. The red spectrum corresponds to the case of $J\neq0$, where the system parameters satisfy the conditions outlined in Eq.(\ref{equ8}). We can find that the OMIT spectrum of the system on the ES with $J=0$ seems the same shape as the standard OMIT spectrum~\cite{weisOptomechanicallyInducedTransparency2010e}. However, a key distinction is that the OMIT spectrum of the system on the ES is robustness to parameter fluctuations. Specifically, as the system moves across different EPs on the ES, the OMIT spectrum remains unchanged, demonstrating the robustness of the ES against parameter fluctuations. Here, we show the transmission spectra at three different EPs as examples in Fig.~\ref{fig2}(b), where the three EPs lie on the same line, as shown in Fig.~\ref{fig2}(c). To more clearly demonstrate the continuous variation of the transmission spectrum on this line in Fig.~\ref{fig2}(c), we plot the transmission spectrum as a function of detuning ($\omega_p-\omega_0$) and $\gamma_1$ in Fig.~\ref{fig2}(d), where $\gamma_2=f(\gamma_1)$. So far, we have shown the ES-protected transmission spectrum, which is useful for studying the spectral properties of the system at the critical phase transition point.

\begin{figure}[htpb]
	\centering
	\includegraphics[width=0.48\textwidth]{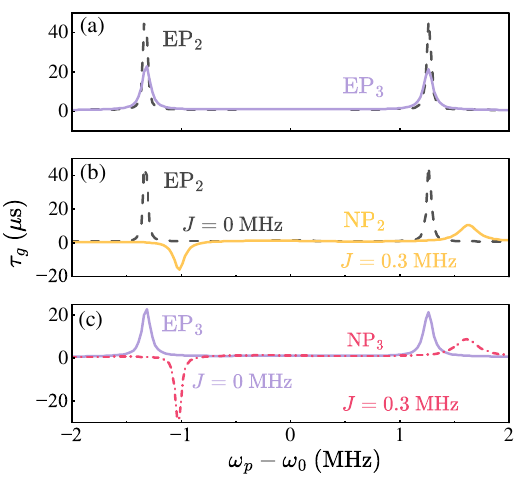}
	\caption{Group delay of the probe light $\tau_{p}$ as a function of the detuning $\omega_p-\omega_0$. (a) The group delay when the system is located at $\rm{EP}_2, \rm{EP}_3$ in Fig.~\ref{fig2}(c). (b,c) The group delay comparison diagrams of EP and it corresponding NP with $J=0.3~\rm{MHz}$.}\label{fig3}
\end{figure}

Moreover, the robustness of the ES against parameter fluctuations is also verified and discussed in fast-slow light effects~\cite{safavi2011electromagnetically,zhou2013slowing}, which could be characterized by optical group delay
\begin{equation}
\tau_{g}=\frac{d\:\rm arg (t)}{d \delta_{p}}.\label{euq16}
\end{equation}
Here, $\tau_g>0$ corresponds to the phenomenon of slow light, and $\tau_g<0$ indicates the fast light appearing. This robustness ensures that fast-slow light effects remain stable across the ES. For example, in Fig.~\ref{fig3}(a), two EPs ($\rm{EP}_2, \rm{EP}_3$) on the same ES both exhibit a slow light effect at the same detuning. The only difference between them is the delay time of the transmission light. However, when the system is taken away from the ES due to the introduction of nanoparticle ($J\neq0$), the fast-slow light effect of the system is completely reversed and the location is also shifted, as shown in Figs.~\ref{fig3} (b) and (c). These results demonstrate that the changes of OMIT and fast-slow light are unobvious for the parameter fluctuations on the ES. However, when the system is taken away from the ES, the nature of the transmission light changes significantly.

\begin{figure*}[htpb]
	\includegraphics[width=0.95\textwidth]{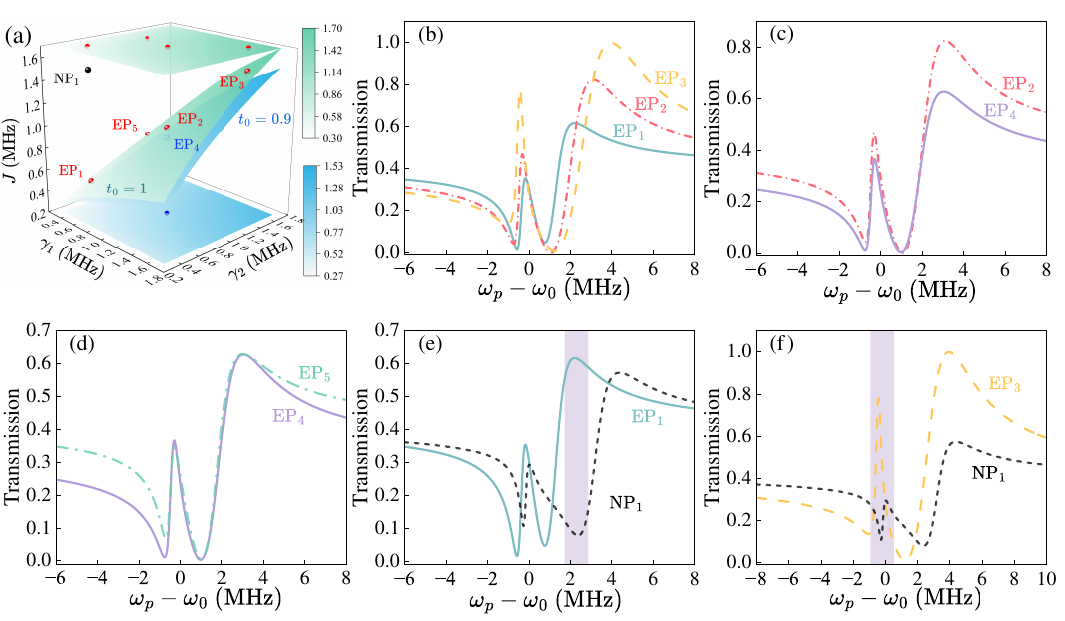}
	\centering
	\caption{(a) Two different ESs with $t_{0}=1$ and $0.9$ of the second kind of ES. Those red and blue points correspond to EPs on different ESs, respectively. And the black point is the normal NP. (b-d) show the comparison between EPs on the same or different ESs, respectively. (e,f) are the comparison between EPs and $\rm{NP}_1$. Please refer to Tab.~\ref{tab1} for the parameters of all points.}\label{fig4}
	\end{figure*}

\subsection{The second kind of exceptional surface}

Here, we consider and discuss the nature of transmission spectra when the system is on the second kind of ES with $J\neq0$. Different from the first kind of ES with $J=0$, the presence of $J$ enforces unidirectional coupling, which induces asymmetric distortions in the spectrum and creates a transparency window at blue detuning, as shown the red line in Fig.~\ref{fig2}(a). It is important to note that Eq.(\ref{equ8}) involves four independent parameters, indicating that the complete ES constitutes a hypersurface within a four-dimensional parameter space. To investigate the properties of EPs on the ES, we have selected two representative cross-sections of the full ES (at $t_0=1$ and $t_0=0.9$), as shown in Fig.~\ref{fig4}(a). Additionally, we have selected five EPs on these two ESs to explore the variation of transmission spectrum of the system within the same ESs or different ESs. And the main parameters of the different EPs are listed in Tab~\ref{tab1}. In Fig.~\ref{fig4}(b), we can see that the OMIT spectra for three EPs located on the same surface exhibit similar trends, which also demonstrate the robustness of the second kind of ES to parameter fluctuations in the OMIT spectrum. Compared to $\rm{EP}_1$ and $\rm{EP}_2$, $\rm{EP}_3$ has a higher dissipation coupling, resulting in a larger effective pump and a greater energy level splitting. This effect is reflected in the OMIT spectrum as a broadening and a shifting of the transparency window to the right. When the system is on different ESs, the transmission window does not undergo the frequency shifting, as shown in Fig.~\ref{fig4}(c). That is because there is only a difference in the transmission coefficient $t_0$ between the two EPs ($\rm{EP}_2$ and $\rm{EP}_4$) on different ESs. The different transmission coefficients only bring about the change in the height and do not affect the position of the transmission window.

On the other hand, we also find that even on different ESs, we can find two points with similar transmission spectral properties, such as $\rm{EP}_4$ and $\rm{EP}_5$ in Fig.~\ref{fig4}(d). This is helpful for arbitrary manipulation of the transmission spectrum of the system under experimental conditions. At last, one NP detached from the two ESs is selected to highlight the change of the OMIT spectrum and the results are given in Fig.~\ref{fig4}(e) and (f). When the system is taken away from the second kind of ES and reaches $\rm{NP}_1$, we can observe that the shape of transmission spectrum undergoes a significant distortion compared with the other EPs. More interesting, we can see that the transparency window induced by the $\rm{EP}_1$ transitions into an absorption valley, as the highlighted zone shown in Fig.~\ref{fig4}(e). We also compare the cases of $\rm{NP}_1$ and $\rm{EP}_3$ in Fig.~\ref{fig4}(f), where the coupling strength $J$ is consistent. We find that the phenomenon of peak-to-valley transition occurs near the resonance $\omega_p-\omega_0\sim0$. Those indicate that, when the system is located at $\rm{NP}_1$, the transmission properties of the system have changed substantially.
\begin{table}[htpb]
  \centering
  \caption{The data of $\rm{NP}_1$ and EPs on ES for $J\neq0$}\label{tab1}
  \begin{tabular}{l||c|c|c|c}\hline
  \toprule[2pt]
  \textbf{} & \textbf{$J$} & \textbf{$\gamma_1$} & \textbf{$\gamma_2$} & \textbf{$t_0$} \\ \hline\hline \noalign{\smallskip}
  \textbf{$\rm NP_1$} & 1.5 MHz & 0.5 MHz & 0.5 MHz & ~~~1~~~ \\ \hline
  \textbf{$\rm EP_1$} & 0.5 MHz & 0.5 MHz & 0.5 MHz & ~~~1~~~  \\ \hline
  \textbf{$\rm EP_2$} & 1 MHz & 1 MHz & 1 MHz & ~~~1~~~  \\ \hline
  \textbf{$\rm EP_3$} & 1.5 MHz & 1.5 MHz & 1.5 MHz & ~~~1~~~  \\ \hline
  \textbf{$\rm EP_4$} & 0.9 MHz & 1 MHz & 1 MHz & ~~~0.9~~~ \\ \hline
  \textbf{$\rm EP_5$} & 0.82 MHz & 0.61 MHz & 1.11 MHz & ~~~1~~~ \\ \hline
\bottomrule[1pt]
\end{tabular}
\end{table}

\begin{figure}[htpb]
\centering
\includegraphics[width=0.48\textwidth]{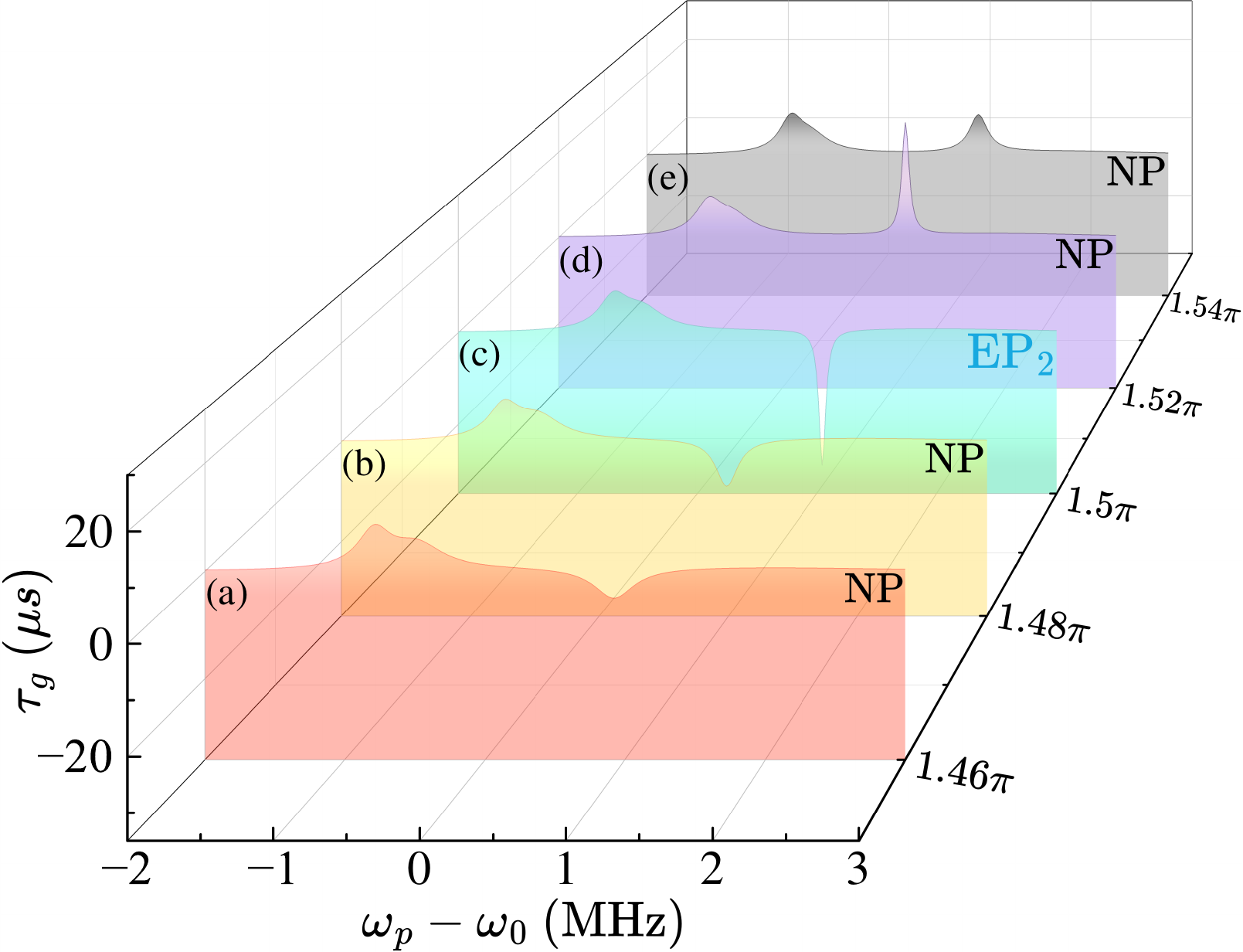}
\caption{Group delay of the probe light $\tau_{g}$ as a function of the detuning
 $\omega_p-\omega_0$. (a, b, d, e) show the group delay of different NPs. (c) The group delay of $\rm{EP_2}$ which is located on the second kind of ES with $J\neq0$.}\label{fig5}
\end{figure}

Finally, we discuss the fast-slow light phenomena when the system is located on the second kind of ES. Here, we can control the phase $\phi_{3}$ to determine whether the system is on the ES, where $\phi_{3}=1.5\pi$ represents the system being on the ES. The detailed results are given in Fig.~\ref{fig5}. We can see that the reversal effect similar to that on the first kind of ES also appears in the fast-slow light phenomena. However, different from the first kind of ES, the system located on the second kind of ES has both fast and slow light phenomena with different detuning, as shown in Fig.~\ref{fig5}(c). When the parameter relationship is broken by changing the phase $\phi_{3}$, the system would take away the second kind of ES. And those results indicate that the fast light effect located around $\omega_p-\omega_0=1~\mathrm{MHz}$ becomes weaker with the phase $\phi_{3}$ decreasing. But when the phase $\phi_{3}$ increases, the fast-slow effect is reversed suddenly. The results not only indirectly demonstrate the robustness of the ES to parameter variations, but also highlight another important advantage of the ES for its high sensitivity when the system takes away the ES. Therefore, the transmission spectra on the ES maybe have significant potential in the field of quantum precision measurement.

Through the analysis and discussion of OMIT and fast-slow light phenomena on two kinds of ESs, complemented by comparisons with prior work~\cite{luOptomechanicallyInducedTransparency2018b}, we observe that all EPs (whether constituting an ES or existing in isolation) induce substantial modifications in OMIT spectra and fast-slow light phenomena when the system deviates from the EP condition. Notably, this behavior represents a defining feature of EPs, manifesting their unique dynamics in non-Hermitian systems. Importantly, the ES exhibits a fundamental advantage over isolated EPs through its formation of a continuous manifold in parameter space. This continuity grants the ES enhanced robustness against parameter variations, meaning that parameters can fluctuate within a certain range without disrupting the EP's ability to control OMIT and fast-slow light effects. Specifically, whereas isolated EPs require exact parameter matching, the ES comprises a continuous collection of EPs, ensuring stable system operation across extended parameter ranges. This inherent robustness significantly improves system reliability and performance, while providing exceptional flexibility and resilience for experimental implementations.

\section{Experimental feasibility}\label{sec4}
In this section, the focus is primarily on discussing the feasibility of theoretical models in experiments. 
The cavity optomechanical system of the WGM type has been extensively studied experimentally
~\cite{peng2016chiral,chenExceptionalPointsEnhance2017c,he2011detecting,peng2014parity,ozdemir2014highly}. 
Therefore, the theoretical model consists of three key components that need to be experimentally realized.
The first part involves the coupling between the WGM transmission mode and an external nanoparticle. 
The second part concerns the dissipative coupling between a tapered fiber and the WGM cavity.  
The third part deals with controlling the transmission rate and phase of light. Based on these experimental and theoretical studies
~\cite{peng2014parity,luOptomechanicallyInducedTransparency2018b,qinExperimentalRealizationSensitivity2021d}, 
we have chosen the feasible parameters according to those WGM experiments. The chosen parameters are 
$R=34.5~\mathrm{\mu m}$, $\omega_{0}=193~\mathrm{THz}$, $\gamma_{0}=1~\mathrm{MHz}$, $m=50~\mathrm{ng}$, 
$\omega_{m}=147~\mathrm{MHz}$, $P_{c}=1~\mathrm{mW}$, $\gamma_{m}=0.24~\mathrm{MHz}$ throughout the discussions.

The two optical modes in the non-Hermitian system exhibit symmetric coupling mediated by a nanoparticle of polarizability $\alpha$ located at position $ \mathbf{r}$. The coupling strength~\cite{PhysRevLett.99.173603} is expressed as
\begin{equation}
	2J = -\alpha f^2(\mathbf{r}) \frac{\omega_0}{V_m},
\end{equation}
where $V_m = \int \rho \, \mathrm{d}x \mathrm{d}y \mathrm{d}z / \max\{\rho\} $ is the mode volume, with $\rho$ representing the optical energy density distribution. $f(\mathbf{r})$ is the field distribution of the WGM, which is determined by the nanoparticle's position, decaying as the nanoparticle separation increases. Therefore, $J$ can be precisely adjusted by using a nanopositioner to control the distance between the nanoparticle and the cavity. Alternatively, $J$ can be tuned by selecting nanoparticles with different materials to modify the polarizability $\alpha$.

The dissipative coupling~\cite{zou2012whispering} between the tapered fiber and the WGM cavity can be described as
\begin{equation}
\gamma_{1,2} = \frac{\eta}{2\tau_c},
\end{equation}
where $\tau_c = 2n\pi R / c$ is the round-trip time of light in the WGM cavity. $\eta$ is proportional to the overlap between the electric field distributions of the tapered fiber and the WGM cavity
\begin{equation}
	\eta \propto \int \mathbf{E}_f \cdot \mathbf{E}_{\text{wgm}}^* e^{-i\Delta \beta z} \mathrm{d}x \mathrm{d}y \mathrm{d}z.
\end{equation}
Here, $\mathbf{E}_f $ and $ \mathbf{E}_{\text{wgm}} $ represent the electric field distributions of the tapered fiber and the WGM cavity, respectively. $ \Delta\beta $ is the difference in the propagation constants of the traveling waves in the tapered fiber and the WGM cavity. For a given fiber and WGM cavity, the propagation constants are fixed. Thus, the simplest way to adjust the coupling coefficient is by varying the distance between the tapered fiber and the WGM cavity, with larger separations yielding weaker coupling. Additionally, we have summarized experimentally achievable parameter ranges for the system configuration in Tab.~\ref{tab2}.
\begin{table}[htpb]
	\centering
	\caption{Summary of Parameters $\gamma_{1,2}$ and $J$}\label{tab2}
	\begin{tabular}{c||c|c}\hline
	\toprule[2pt]
	 &$\gamma_{1,2}$ &   $J$  \\ \hline\hline \noalign{\smallskip}
	 Ref.~\cite{wang2020electromagnetically} & 5.57-11.98~MHz &  0.22-7.11~MHz \\ \hline
	 Ref.~\cite{pengLossinducedSuppressionRevival2014b} & - & 0-200~MHz \\ \hline
	 Ref.~\cite{qinExperimentalRealizationSensitivity2021d}& 0.1-3~MHz & 0.87~MHz \\ \hline
     Ref.~\cite{PhysRevLett.118.033901} & 0.87-5.84~MHz & - \\ \hline
  \bottomrule[1pt]
\end{tabular}
\end{table}

The unidirectional coupling can be achieved by adding an isolator at the end of the waveguide in experiments\cite{jiangExperimentalRealizationExceptional2023a,qinExperimentalRealizationSensitivity2021d,soleymani2022chiral}. Furthermore, the phase of light fields can be adjusted by the PM. The transmission coefficient $t_{0}$ of the waveguide can be changed by changing the waveguide length in actual experiment~\cite{qinExperimentalRealizationSensitivity2021d}. Anyway, the experimental parameters and techniques involved in our protocol are fully feasible under the current experimental conditions.

\section{CONCLUSION}\label{sec5}
In conclusion, we present a novel scheme aimed at investigating the behavior of OMIT on two kinds of ES within a non-Hermitian WGM-type optomechanical system. The inherent non-Hermitian characteristics of the system stem from the intricately designed waveguide structure. In comparison to a single EP, ES not only offers an infinite number of EPs for exploring the impact of phase transition points on the OMIT but also displays remarkable robustness to parameter variation, which is friendly for experimental design~\cite{zhongSensingExceptionalSurfaces2019c,qinExperimentalRealizationSensitivity2021d}. Our investigation delves into two kinds of ESs, revealing their robustness to parameter variations in OMIT spectra and fast-slow light effects. Moreover, by introducing a nanoparticle or modulating the phase, we disrupt the ES conditions, thereby transitioning the system to the NP case. It is notable that deviations from the ES lead to a loss of robustness in the transparency window of the OMIT spectrum, potentially resulting in the peak-valley conversions. Furthermore, we delve into the fast-slow effects observed in transmission spectra across various ES configurations. Our findings underscore the potential conversion of fast-slow phenomena when the system deviates from the ESs. This comprehensive exploration not only enriches the understanding of information transmission in non-Hermitian quantum systems but also offers valuable insights that enhance the flexibility in parameter selection for experimental setups. Ultimately, our research would pave the way for the realization of ES-enhanced robust quantum sensing applications, offering a theoretical framework that could prove fundamental for future advancements in this field.

\begin{acknowledgments} 
This work was supported by the National Natural Science Foundation of China (Grants No. 12204424, No. 12147149, No. 12405029, No. 12205256, No. 12274376, No. 11935006, No. 12421005), the China Postdoctoral Science Foundation (Grant No. 2022M722889), the Sci-Tech Innovation Program of Hunan Province (Grant No. 2020RC4047), the National Key R\&D Program (Grant No. 2024YFE0102400), the Hunan Major Sci-Tech Program (Grant No. 2023ZJ1010), the Henan Science and Technology Major Project of the Department of Science and Technology of Henan Province (Grant No. 241100210400), the Natural Science Foundation of Henan Province (Grant No. 242300420665), and the Doctoral Research Foundation of Zhengzhou University of Light Industry (Grant No. 2022BSJJZK20).
\end{acknowledgments}

\appendix
\section{The detailed results of $\delta a_{\rm{cw}}^-$ and $\delta a_{\rm{ccw}}^-$}\label{sec7}
In Eq.~(\ref{equ13}), we have five quintuple first order equations. By solving these five equations, the detailed expression of $\delta a_{\rm{cw}}^-$ and $\delta a_{\rm{ccw}}^-$ can be calculated, individually. The results are obtained by
\begin{align}
\delta a_{\rm{cw}}^{-}=E_p*(A_1+iA_2)/B,\nonumber\\
\delta a_{\rm{ccw}}^{-}=E_p*(A_3+iA_4)/B.
\end{align}
The detailed expressions of $A_1,~A_2,~A_3,~A_4$, and $B$ are as follows
\begin{align}
 A_1&=d_1\sqrt{\gamma_2}*{\rm Re}(t_2)+\sqrt{\gamma_1}[d_2-i
  h_2h_7],\nonumber\\
  A_2&=d_1\sqrt{\gamma_2}*{\rm Im}(t_2),\nonumber\\
  A_3&=\sqrt{\gamma_1}h_7(k_1*{\rm Re}(t_2)+ih_4)-i\sqrt{\gamma_1}h_6d_2/f_1\nonumber\\
  &~~+f_1\sqrt{\gamma_2}(if_2h_{12}+ih_3h_6^*-h_4h_6^*)*{\rm Re}(t_2),\nonumber\\
  A_4&=\sqrt{\gamma_2}(d_2-ih_1h_7)*{\rm Im}(t_2),\nonumber\\
B&=t_0^4\gamma_1^2\gamma_2^2+i\sqrt{\gamma_1\gamma_2}h_{43}(t_3^*k_2+t_3J^{*2})\nonumber\\
  &~~+\gamma_1\gamma_2[it_0^2\sqrt{\gamma_1\gamma_2}(t_3+t_3^*)h_{43}\nonumber\\
  &~~-(Jt_3^{*2}+J^*t_3^2)h_{34}+t_3^{*2}k_2+t_3^2J^{*2}-it_3^{*2}f_1h_{12}]\nonumber\\
  &~~+J^*[k_2J^*-if_1J^*h_{12}-(k_2+JJ^*)h_{34}],
\end{align}
where
\begin{align}
d_1 &= -t_3t_3^{*2}(\gamma_1\gamma_2)^{3/2}+ f_2h_5h_{12} \nonumber\\
&~~+ \sqrt{\gamma_1\gamma_2}[t_3J^*(h_{34}-J^*) + t_3^{*}Jh_{43}]\nonumber\\
&~~-i \big[f_2^2h_{53} + t_3^{*}\gamma_1\gamma_2((J-h_3)t_3^* + t_3h_{43})\nonumber\\
&~~- J^*(Jh_{34} + J^*h_3 - J^2)\big], \nonumber\\
d_2 &= f_1\big[h_7 + i(t_3^*\sqrt{\gamma_1\gamma_2}h_4 + f_2h_{12}) - h_3h_5^* - h_4J^*\big], \nonumber\\
h_1 &= g^2\chi \overline{a}_{\rm{cw}}\overline{a}_{\rm{cw}}^{*}, \nonumber\\
h_2 &= g^2\chi \overline{a}_{\rm{ccw}}\overline{a}_{\rm{ccw}}^{*}, \nonumber\\
h_3 &= g^2\chi \overline{a}_{\rm{cw}}\overline{a}_{\rm{ccw}}^{*}, \nonumber\\
h_4 &= g^2\chi \overline{a}_{\rm{ccw}}\overline{a}_{\rm{cw}}^{*}, \nonumber\\
h_5 &= J + it_3\sqrt{\gamma_1\gamma_2}, \nonumber\\
h_6 &= J - it_3\sqrt{\gamma_1\gamma_2}, \nonumber\\
h_7 &= f_2^2 + J^{*2} + \gamma_1\gamma_2t_3^{*2}, \nonumber\\
k_1 &= f_1 - ih_1, \nonumber\\
k_2 &= J^2 + f_1^2, \nonumber\\
h_{nl} &= 
\begin{cases}
h_{n} + h_{l} & \text{if } n < l, \\
h_{n} - h_{l} & \text{if } n > l.
\end{cases}
\end{align}

\bibliography{jab111}

\end{document}